\documentclass[twocolumn,preprintnumbers,amsmath,amssymb]{revtex4-1}

\usepackage{graphicx}
\usepackage{dcolumn}
\usepackage{bm}
\usepackage{amsmath}

\newcommand{\rhat}{\hat{\mathbf{r}}}
\newcommand{\op}[1]{\hat{\boldsymbol{#1}}}

\newcommand{\mA}{\mathcal{A}}

\newcommand{\mG}{\mathcal{G}}

\newcommand{\mM}{\mathcal{M}}

\newcommand{\mO}{\mathcal{O}}

\newcommand{\bk}{\mathbf{k}}

\newcommand{\br}{\mathbf{r}}

\newcommand{\btau}{\boldsymbol{\tau}}

\newcommand{\tS}{\text{S}}
\newcommand{\tU}{\text{U}}

\newcommand{\Nmax}{N_{\text{max}}}
\DeclareMathOperator*{\Tr}{Tr}

\begin{document}
\title{Charge self-consistent dynamical mean-field theory based on the full-potential linear muffin-tin orbital method: methodology and applications.}
\author{Oscar Gr\aa n\"as}
\author{Igor di Marco}
\author{Patrik Thunstr\"om} 
\author{Lars Nordstr\"om}
\author{Olle Eriksson}
 \affiliation{Department of Physics and Astronomy, Uppsala University, Box 516, 751 20 Uppsala, Sweden}

\author{Torbj\"orn Bj\"orkman}
 \affiliation{\mbox{COMP, Department of Applied Physics, Aalto University School of Science},\\ P.O.Box 11100, FI-00076 AALTO, Finland}

\author{J. M. Wills}
  \affiliation{Los Alamos National Laboratory, Los Alamos, New Mexico 87545, USA}

\date{\today}

\begin{abstract}
Full charge self-consistence (CSC) over the electron density has been implemented into the local density approximation plus dynamical mean-field theory (LDA+DMFT) scheme based on a full-potential linear muffin-tin orbital method (FP-LMTO). Computational details on the construction of the electron density from the density matrix are provided. The method is tested on the prototypical charge-transfer insulator NiO using a simple static Hartree-Fock approximation as impurity solver. The spectral and ground state properties of bcc Fe are then addressed, by means of the spin-polarized T-matrix fluctuation exchange solver (SPTF). Finally the permanent magnet SmCo$_5$ is studied using multiple impurity solvers, SPTF and Hubbard I, as the strength of the local Coulomb interaction on the Sm and Co sites are drastically different. The developed CSC-DMFT method is shown to in general improve on materials
properties like magnetic moments, electronic structure and the materials density.

\end{abstract}
\maketitle

\section{Introduction}
Understanding the interplay between strong electron-electron interactions and kinematic effects in crystals pose a major experimental and theoretical challenge. Although the problem is very demanding, the potential rewards are great. Many interesting materials properties, like high temperature superconductivity\cite{bednorz86ZpB64:189} and coexistence of superconductivity and magnetism, are found in materials with a correlated electronic structure,  e.g. the newly discovered Fe-pnictides\cite{kamihara08jacs130:3296}. Strong correlations are also associated with mixed valence and heavy Fermion materials.

On the theoretical side it now stands clear that a straightforward application of conventional electronic structure theory\cite{kohn65pr140:A1133,hohenberg64pr136:B864}, as given by the local density approximation (LDA) or generalized gradient approximation (GGA), often fails in describing correlated electron materials. In addition it is evident that an approach based on model Hamiltonians also has drawbacks in that many of the parameters of these models are hard to estimate and that the band-formation often is described inaccurately. Hence it becomes important to combine the best features of the two approaches, something which can be achieved by combining density functional theory in the local density approximation and dynamical mean field theory (LDA+DMFT)\cite{georges96rmp68:13,kotliar06rmp78:865}. To date there have been several reports that integrate DMFT with existing codes for electronic structure calculation. Starting with the pioneering papers of Ref.\onlinecite{savrasov01nature410:793,anisimov97jpcm9:7359,pavarini04prl92:176403,lichtenstein98prb57:6884}, various implementations have been produced, differing in the parent electronic structure method and the way the local orbitals are constructed. At the beginning most of such implementations were limited to work with a fixed LDA electron density. This was not only due to the technical difficulties in constructing a reliable electron density from the DMFT Green's functions, but also to the fact that fully charge self-consistent (CSC) simulations would have needed a prohibitive amount of computational resources. However, in the last years these limitations have been overtaken by following two different approaches. A straight forward but approximate computational scheme has been proposed by associating the new electron density to an LDA+U problem \cite{anisimov97jpcm9:767} constructed with the DMFT density matrix \cite{shick09prb80:085106,suzuki09prb80:161103}. Several fully self-consistent LDA+DMFT suites have also been produced \cite{savrasov04prb69:245101,minar05prb72:045125,pourovskii07prb76:235101,aichhorn09prb80:085101,haule10prb81:195107}, although it remains to be seen if their applicability is limited by the high computational cost.
 
In the present paper we report the extension of an LDA+DMFT implementation\cite{grechnev07prb76:035107,thunstrom09prb79:165104,dimarco09prb79:115111} based on the full-potential linear muffin-tin orbitals (LMTO) method\cite{wills87prb36:3809,wills:fp-lmto,RSPt_book,rspt_website} in which we include full self-consistency in the electronic density. Due to the full potential character such an implementation is suitable for simulations of any system, without restriction on the geometry. In addition the usage of a small number of basis functions associated to the LMTO method makes it accessible to calculate the fully hybridizing electronic structure of very large systems without any ad-hoc assumptions. The technical details of the current development, together with a brief review of the DMFT equations, are presented in Section II, and in Appendix A.

In section III, we apply the developed LDA+DMFT scheme to several materials with various degrees of electron correlation, to illustrate the usefulness of the method. Our first test-case is the prototypical insulator NiO in its sodium-chloride phase. This is an archetypical compound which is often used to illustrate that electronic structures based on LDA or GGA alone fail\cite{anisimov91prb44:943}. Our second test case is represented by the itinerant ferromagnet bcc Fe. Here the effects of electron correlations are much less pronounced, but several well known failures with conventional LDA/GGA theory are known, e.g. the too broad band widths of the 3d states\cite{barriga09prl103:267203} and the presence of a weak intermediate-energy satellite\cite{grechnev07prb76:035107}, which is observed in photoemission experiments\cite{hufner00prb61:12582}. Finally we study the electronic structure of the rare-earth based permanent magnet SmCo$_5$, which is a particularly useful case, since the correlated electrons on the Sm and Co sites experience different strenght of the local Coulomb interaction, and therefore require different impurity solvers.

In the last section of this paper, section IV, we present our conclusions and perspectives of further development and future applications.

\section{The charge self-consistent LDA+DMFT scheme}
The starting point of the LDA+DMFT scheme is the assumption that the single-particle LDA Hamiltonian $\op{H}^{LDA}$ does not correctly describe the exchange and correlation effects associated to a certain set of local orbitals $\left\{|R,\xi\rangle\right\}$, where $R$ denotes the lattice site and $\xi$ the quantum numbers labeling such states. A local two-particle operator $\op{U}_R$, describing an effective Coulomb interaction between the Kohn-Sham quasi-particles, is therefore added to $\op{H}^{LDA}$, giving the so-called LDA+U Hamiltonian \cite{aguilar84pss123:219,thalmeier79prb20:4637,anisimov97jpcm9:767}:
\begin{equation}
\label{eq:ldauh}
 \op{H}^{LDA+U}=\op{H}^{LDA} + \sum_R \op{U}_R - \op{H}_R^{DC}.
\end{equation}
From this point on the site subscript $R$ denotes that the (eventually $\bk$-dependent) operator is projected onto the local orbitals
\begin{equation}
\op{A}_R \equiv \sum_{\xi,\xi'} |R,\xi\rangle\langle R,\xi| \sum_\bk \op{A}_\bk |R,\xi'\rangle\langle R,\xi'|.
\end{equation}
The double-counting correction $\op{H}_R^{DC}$ in Equation (\ref{eq:ldauh}) is introduced to remove the improper LDA description of the local Coulomb interaction. It is clear that there are many issues related to the arbitrariness of the choice of the correlated orbitals, and to the difficulty of determining a reliable double-counting correction. However, a discussion of such topics would be out of the scope of this paper, so we refer the interested reader to some recent reviews\cite{georges96rmp68:13,kotliar06rmp78:865} for more details regarding the LDA+U Hamiltonian and the DMFT.

\begin{figure}
\includegraphics[scale=0.7]{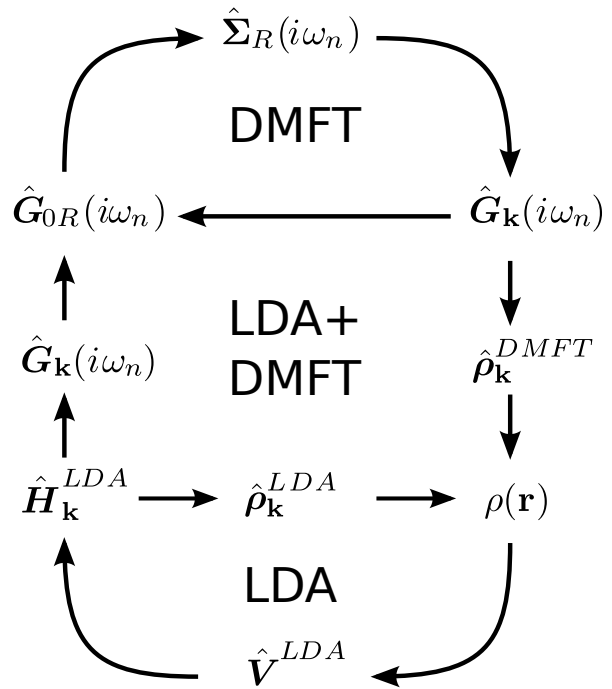}
\caption{The charge self-consistent LDA+DMFT scheme. The LDA+DMFT cycle is the union of the LDA cycle (lower section) and the DMFT cycle (upper section). A charge non-self-consistent (one-shot) LDA+DMFT scheme lacks the construction of $\op{\rho}^{DMFT}_\bk$, which implies that $\op{H}^{LDA}$ is not updated.}
\label{fig:cycles}
\end{figure}

The main idea of the DMFT is that the Hubbard model described by the LDA+U Hamiltonian can be mapped locally to an effective Anderson impurity model (AIM)\cite{kotliar06rmp78:865}. In this way the lattice problem is reduced to the much simpler problem of an atom embedded in a electronic bath. This leads to the LDA+DMFT scheme depicted in Figure \ref{fig:cycles}. It consists of two interconnected self-consistency cycles, the LDA cycle and the DMFT cycle, which are presented in more detail below. 

\subsection{The LDA cycle}
The starting point of the LDA cycle is the calculation of the effective LDA potential $\op{V}^{LDA}$ from some electron density $\rho(\br)$, using the LDA functional. $\op{V}^{LDA}$ is then used to construct a $\bk$-point dependent LDA Hamiltonian $\op{H}_\bk^{LDA}$. The pure LDA cycle can then be closed by constructing a density matrix
\begin{equation}
\op{\rho}_\bk^{LDA} = \left[\op{1}+\exp\left(-\beta(\op{H}^{LDA}_\bk-\mu\op{1})\right)\right]^{-1}, \label{eqn:rholdaan}
\end{equation}
from which a new electron density $\rho(\br)$ can be obtained. In Equation (\ref{eqn:rholdaan}) the chemical potential $\mu$ is set to ensure that the system contains the correct number of electrons at temperature $T=1/\beta$. This pure LDA cycle is shown in the lower part of Fig.\ref{fig:cycles}. After the LDA cycle has produced a reasonable initial electronic structure, one may exit the LDA cycle and continue into the DMFT part of the full LDA+DMFT scheme. Once the calculation has entered the DMFT cycle, the continued use of the LDA cycle will be limited to updating $\op{H}^{LDA}_\bk$ from $\op{\rho}_\bk^{DMFT}$. Details of how to construct the electron density from a representation of the density matrix in the LMTO basis are presented in Appendix \ref{density}.

\subsection{The DMFT cycle} \label{sec:ldadmft_eqs}
The DMFT cycle starts with the creation of the $\bk$-dependent Green's function $\op{G}_\bk(i\omega_n)$ from $\op{H}^{LDA}_\bk$ and a set of local self-energies $\{\op{\Sigma}_R(i\omega_n)\}$
\begin{equation}
\op{G}_\bk(i\omega_n) = \left[i\omega_n\op{1}+\mu\op{1} - \op{H}^{LDA}_\bk - \sum_R\op{\Sigma}_R(i\omega_n)\right]^{-1},\label{eqn:gglobal}
\end{equation}
where $\op{G}_\bk(i\omega_n)$ is resolved over the Matsubara frequencies $i\omega_n = i\pi T (2n+1)$ from the finite temperature many-body formalism. At the first iteration the $\bk$-independent self-energies are set to zero, but in the following ones they will be obtained from the solution of the effective impurity problem. The chemical potential $\mu$ in Equation (\ref{eqn:gglobal}) needs to be readjusted at every DMFT iteration to ensure that the number of electrons in the system is conserved. The mapping to the effective impurity model is done through the use of local bath Green's functions $\op{\mathcal{G}}_{0R}(i\omega_n)$, which are obtained by means of an inverse Dyson equation,
\begin{equation}
\op{\mathcal{G}}_{0R}^{-1}(i\omega_n) = \op{G}^{-1}_R(i\omega_n) + \op{\Sigma}_R(i\omega_n).\label{eqn:dyson}
\end{equation}
The $\op{U}$ term in Equation (\ref{eq:ldauh}) and the bath Green's function $\op{\mathcal{G}}_{0R}(i\omega_n)$ fully describes the effective AIM. This model is well studied in literature and there exists many impurity solvers. The resulting set of impurity self-energies $\op{\Sigma}^{AIM}_R(i\omega_n)$ are then used to update the local self-energies after the double counting contribution has been removed
\begin{equation}
\op{\Sigma}_R(i\omega_n) = \op{\Sigma}^{AIM}_R(i\omega_n) - \op{H}_R^{DC}.
\end{equation}
The DMFT cycle can now be continued using the updated self-energies, as shown in the upper part of Fig. \ref{fig:cycles}. Alternatively, a new density matrix $\op{\rho}^{DMFT}_\bk$ can be calculated from the $\bk$-dependent Green's function, to update $\op{H}^{LDA}_\bk$ in the LDA cycle.

\subsection{Updating the electron density}
The density matrix $\op{\rho}^{DMFT}_\bk$ can be obtained from the Green's function as an infinite sum over all the Matsubara frequencies \cite{fetter-walecka03}
\begin{equation}
\op{\rho}_\bk = \lim_{\eta \rightarrow 0^+} \lim_{N\rightarrow\infty} T \sum^N_{n=0} \left[\op{G}_\bk(i\omega_n)e^{i\omega_n\eta} +  \op{G}^\dagger_\bk(i\omega_n)e^{-i\omega_n\eta}\right].\label{eqn:rho}
\end{equation}
The order of the limits can not be interchanged as the partial sums are only point wise convergent and not uniformly convergent with respect to $\eta$. A way around this problem is to decompose the Green's function into a numerical part and an analytical part, taking advantage of the asymptotic expansion of the self-energy for large $\omega_n$
\begin{equation}
\op{\Sigma}_R(i\omega_n) = \op{\Sigma}_R(\infty) + O\left(\frac{1}{i\omega_n}\right).
\end{equation}
Defining
\begin{eqnarray}
\op{G}^{an}_\bk(i\omega_n) & = & \left[i\omega_n\op{1}+\mu\op{1} - \op{H}^{LDA}_k - \sum_R\op{\Sigma}_R(\infty)\right]^{-1}\\[1ex]
\op{G}^{num}_\bk(i\omega_n) & = & \op{G}_\bk(i\omega_n) - \op{G}^{an}_\bk(i\omega_n),
\end{eqnarray}
the density matrix can be split into two parts as $\op{\rho}_\bk = \op{\rho}^{an}_\bk + \op{\rho}^{num}_\bk$, where
\begin{eqnarray}
 \op{\rho}^{an}_\bk & = & \lim_{\eta \rightarrow 0^+} \lim_{N\rightarrow\infty} T \\
 & & \sum^N_{n=0}  \left[{\op{G}^{an}_\bk(i\omega_n)e^{i\omega_n\eta} +  \op{G}^{an\dagger}_\bk(i\omega_n)e^{-i\omega_n\eta}}\right]  \nonumber\\[2ex]
 \op{\rho}^{num}_\bk & = & \lim_{\eta \rightarrow 0^+} \lim_{N\rightarrow\infty} T \\
 & & \sum^N_{n=0} \left[{\op{G}^{num}_\bk(i\omega_n)e^{i\omega_n\eta} +  \op{G}^{num\dagger}_\bk(i\omega_n)e^{-i\omega_n\eta}}\right].\label{eqn:rhonum}  \nonumber
\end{eqnarray}
The analytical part $\op{\rho}^{an}_\bk$ has a simple form but contains the logarithmic divergence of $\op{\rho}_\bk$, while $\op{\rho}^{num}_\bk$ converges uniformly \cite{dimarco09prb79:115111}.  The uniform convergence allows the order of the limits in Equation (\ref{eqn:rhonum}) to be interchanged. With a minimal loss of accuracy the resulting sum can be truncated at some large cut-off Matsubara frequency $N_{\text{max}}$, giving
\begin{equation} \label{eq:rho_num}
\op{\rho}^{num}_\bk \approx \sum^{\Nmax}_{n=0} \left[{\op{G}^{num}_\bk(i\omega_n) + \op{G}^{num\dagger}_\bk(i\omega_n)}\right].
\end{equation}
The analytical part of the density matrix can still not be summed explicitly, but thanks to its frequency independent form it can be evaluated in the same way as $\op{\rho}^{LDA}_\bk$ in Equation (\ref{eqn:rholdaan}), which yields
\begin{equation}
\op{\rho}^{an}_\bk = \left[\op{1}+\exp\left(-\beta(\op{H}^{LDA}_\bk + \sum_R\op{\Sigma}_R(\infty)-\mu\op{1})\right)\right]^{-1}.
\end{equation}
Notice that in the absence of a self-energy $\op{G}^{num}_\bk(i\omega_n) = 0$, and $\op{\rho}^{an}_\bk$ is reduced to $\op{\rho}^{LDA}_\bk$.

\subsection{The asymptotic limit of $\op{\Sigma}_R(i\omega_n)$}
The problem of evaluating the infinite sum in Equation (\ref{eqn:rho}) has now been replaced by finding the value of the Hartree-Fock-like self-energy $\op{\Sigma}_R(\infty)$ from the asymptotic limit of $\op{\Sigma}_R(i\omega_n)$. The asymptotic behavior of $\op{\Sigma}_R(i\omega_n)$ is dictated by the Dyson equation in Equation (\ref{eqn:dyson}). Unless $\Nmax$ is set to a very large value, at a high cost of computational resources, the slow decay of the anti-hermitian part of $\op{\Sigma}_R(i\omega_n)$ makes the zeroth order approximation $\op{\Sigma}_R(\infty) \approx \op{\Sigma}_R(i\omega_{\Nmax})$ far too crude. However, the situation can be improved thanks to the hermiticity of $\op{\Sigma}_R(\infty)$. The contribution from the slowly decaying tails can be canceled out by taking 
\begin{equation}
\op{\Sigma}_R(\infty) \approx \op{\Sigma}^H_R(i\omega_{\Nmax}) \equiv \frac{\op{\Sigma}_R(i\omega_{\Nmax}) + \op{\Sigma}^\dagger_R(i\omega_{\Nmax})}{2}. 
\end{equation}

The current implementation goes one step further and extrapolates $\op{\Sigma}^H_R(i\omega_n)$ to infinity from a least squares fit of a few data points around $\Nmax$. Since $\op{\Sigma}^H_R(i\omega_n)$ is hermitian it can be expanded in orders of $\omega_n^{-2}$, so the first order fit has an error proportional to $\omega_{\Nmax}^{-4}$.

\section{Results}
The following calculations were performed with an LMTO basis set containing a triple basis for s and p states and a double basis for d and f states, where the multiplicity is referred to the number of envelope functions (tails). The basis of the valence electrons was constructed considering 4s, 4p and 3d states for the transition metals atoms, 2s, 2p and 3d states for the O atoms, and 6s, 6p, 5d, and 4f states for the Sm atoms. The correlated basis functions were chosen to be the MT orbitals described in Ref. \onlinecite{IgorThesis}. The {\bf{k}}-points were distributed according to the conventional Monkhorst-Pack grid, and the Brillouin zone integration was carried out using Fermi smearing for the temperature T = 400 K. Finally the U-matrix in Equation (\ref{eq:ldauh}) was constructed from the Slater parameters $F^0$, $F^2$, $F^4$, and $F^6$ \cite{kotliar06rmp78:865}. For Fe and Co an additional parametrization has been adopted by using fixed atomic ratios for $F^2$ and $F^4$, which allows to refer to the bare Coulomb repulsion $U=F^0$ and to the Hund's exchange $J$\cite{anisimov93prb48:16929}.  

\subsection{Nickel oxide}
The first test of our CSC LDA+DMFT implementation is the study of the antiferromagnetic (AFM) NiO within the LDA+U approximation. Although the LDA+U scheme does not necessarily require the usage of a Green's function formalism, it can still be rewritten in terms of the equations presented in Section \ref{sec:ldadmft_eqs}, solving the effective impurity model by means of the Hartree-Fock approximation. NiO in LDA+U is a good test since it is the standard test-case used in most LDA+U implementations and much material is available in the literature. Moreover the resulting self-energy is static, and therefore calculations can be made using only a single Matsubara frequency. This implies that the numerical contribution $\op{\rho}^{num}_\bk$ is zero by construction. Finally it is important to stress that a proper description of the electronic properties of NiO would require a more sophisticated solver, especially if a detailed comparison with experimental data, e.g. from photoemission spectroscopy, is under focus\cite{ren06prb74:195114,kunes07prl99:156404,kunes07prb75:165115}.

The crystal structure of NiO is NaCl but due to the antiferromagnetic structure, the calculations involved 4 atoms per unit cell, 2 O and 2 Ni. The lattice constant used was 7.89 a.u. and the Brillouin zone was sampled through a mesh of 9 x 9 x 9 $\bk$-points, giving 365 points in the irreducible wedge. Spin-orbit coupling was considered with spin quantized along the easy axis (111).  The values $F_0=8.00 \text{ eV}$, $F_2=8.19 \text{ eV}$ and $F_4=5.11 \text{ eV}$ were taken from Ref. \onlinecite{shick99prb60:10763}. Finally the double-counting correction was chosen to be the fully localized limit \cite{IgorThesis}, as is usual for insulators.

\begin{figure}[ht]
\centering
\includegraphics[scale=0.3]{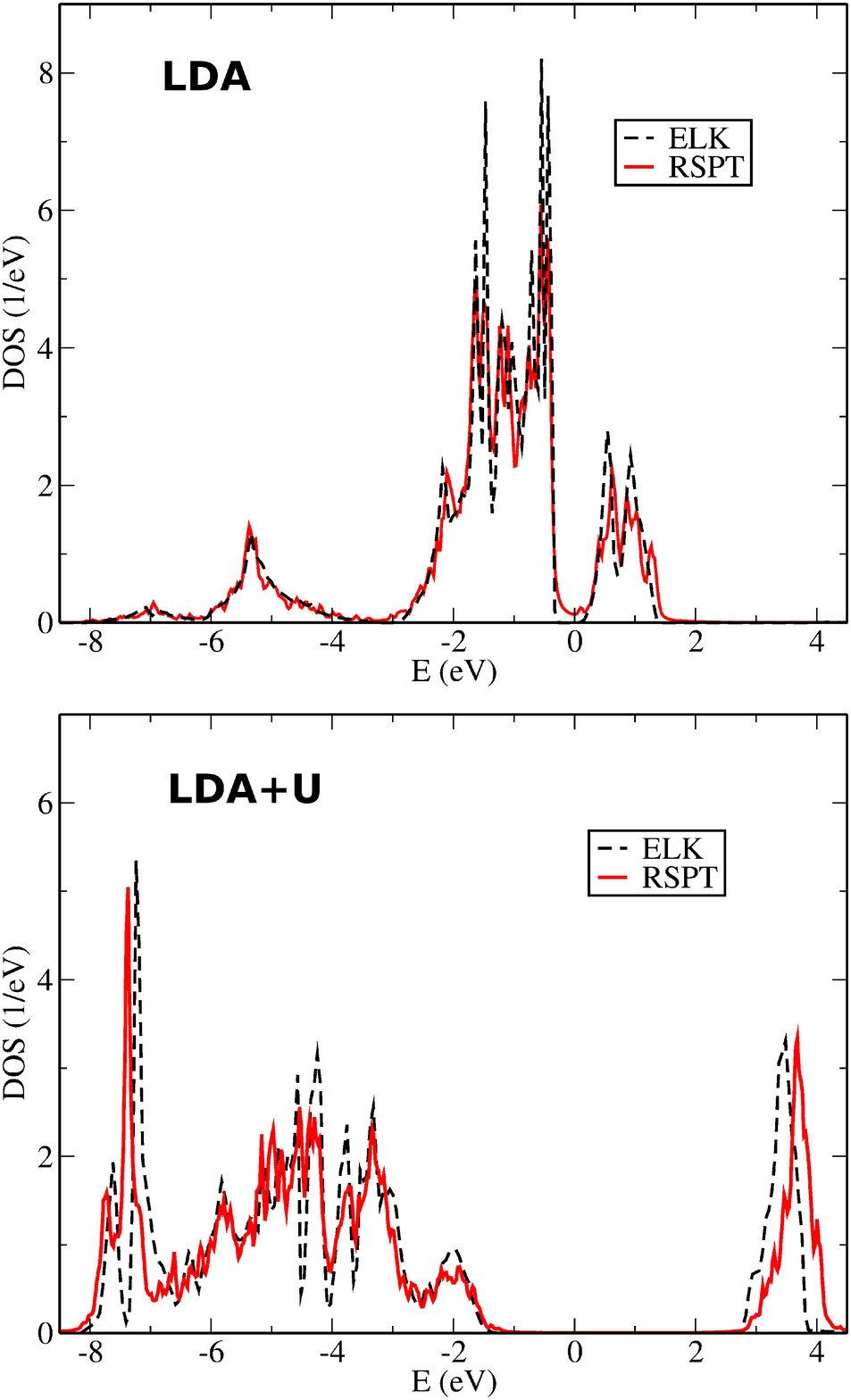}
\vspace{0.2cm}
\caption{(Color online) LDA and LDA+U spin-integrated 3d density of states of NiO in the present RSPt implementation and in the ELK code\cite{elk_website}. The computational details are described in the main text. Notice that the curves have been shifted in order to allign the first occupied peak, due to the arbitrariness of the chemical potential in insulators.}
\label{fig:NiO}
\end{figure}

\begin{table*}
\caption{\label{tab:NiO}Ni 3d contribution to the spin moment, m$_\text{s}$, orbital moment, m$_\text{o}$, and total moment, m$_\text{tot}$, for NiO in RSPt and ELK. These values are compared with the experimental data from Ref. \onlinecite{neubeck99jap85:4847,fernandez98prb57:7870}. All the magnetic moments are given in units of $\mu_B$ per atom.} 
\vspace{0.2cm}
\begin{tabular}{| l | c c c | c c c |}
\hline
 & \multicolumn{3}{c|}{RSPt} & \multicolumn{3}{c|}{ELK} \\
Method & m$_\text{s}$ & m$_\text{o}$ & m$_\text{tot}$ & m$_\text{s}$ & m$_\text{o}$ & m$_\text{tot}$ \\
\hline
LDA & 1.21 & 0.14  & 1.35 & 1.14  & 0.14 & 1.28 \\
LDA+U (CSC) & 1.73 & 0.24  & 1.97 & 1.70 & 0.26  & 1.96 \\
Exp. \cite{neubeck99jap85:4847,fernandez98prb57:7870} & $1.7 \div 2.1$ & $0.28 \div 0.37$ & $2.0 \div 2.4$ & $1.7 \div 2.1$ & $0.28 \div 0.37$ & $2.0 \div 2.4$  \\
\hline
\end{tabular}
\end{table*}

The electronic structure is shown in Figure \ref{fig:NiO} for both the LDA and the LDA+U approximations. The figure contains a comparison between results of the present implementation and the ELK code \cite{elk_website}, which is based on the full-potential linear augmented plane-wave (FP-LAPW) method. As is clear from the figure, the electronic structures of both methods are practically identical when LDA is considered. 
When LDA+U corrections are introduced the main features of the spectral properties are also in good agreement, but some smaller differences may be found for the finer details. These discrepancies can be linked to a different definition of correlated orbitals, which leads to a slightly stronger effect of the Hubbard U for the RSPt results in comparison with the ELK calculations. This can be seen in the peak structure at -7.5 eV, which is found to lie a bit lower in the RSPt implementation, and the unoccupied peak at 3.5 eV, which is a bit higher. For both implementations the electron-electron interaction, parametrized in the form of U, widens the band-gap, and the overall agreement of these results with previous LMTO-based studies is excellent \cite{dudarev98prb57:1505,shick99prb60:10763}.

Some additional information can be gathered by looking at magnetic moments, shown in Table \ref{tab:NiO}. We first notice that the agreement between RSPt results and ELK calculations is very good, and both methods give results compatible with recent experimental data\cite{neubeck99jap85:4847,fernandez98prb57:7870}. 


\subsection{bcc Fe}

We turn next our attention to a system where the electron-electron interaction among the 3d states is weaker than that of NiO, namely bcc Fe. In general for the itinerant ferromagnets a static solver as the Hartree-Fock method used in the previous section is not sufficient, since the energy dependence of the self-energy plays a relatively important role. Instead we have for these calculations used the more sophisticated SPTF solver \cite{katsnelson02epjb30:9,pourovskii05prb72:115106}. Since correlation effects are weak, we do not expect big changes associated to the CSC cycle in the LDA+DMFT scheme. However the self-energy is now a dynamical quantity, and therefore the numerical contribution to the electron density $\op{\rho}^{num}_\bk$ becomes finite. 

The spectral properties of bcc Fe were calculated at the experimental lattice constant \cite{grechnev07prb76:035107} and with spin-orbit coupling with the spin axis aligned along the easy axis (001). The Brillouin zone was sampled through a mesh of 21 x 21 x 21 $\bk$-points, leading to 726 points in the irreducible wedge. Finally the averaged static part of the self-energy was used as double-counting correction \cite{IgorThesis}.

\begin{figure}[ht]
\centering
\includegraphics[scale=0.29]{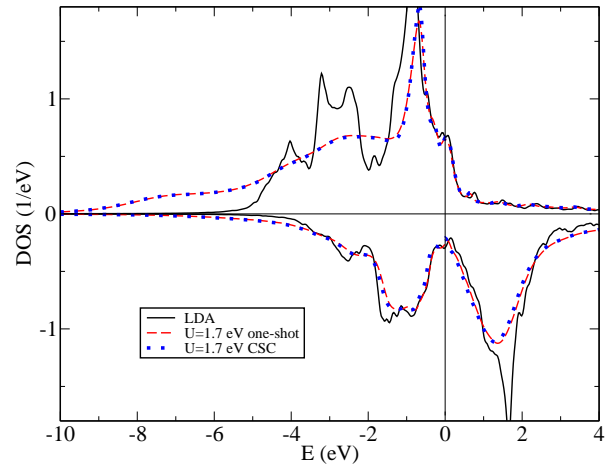}
\vspace{0.2cm}
\caption{(Color online) Projected 3d density of states for bcc Fe in the RSPt code for basic LDA, one-shot LDA+DMFT and CSC LDA+DMFT for U=1.7 eV. The Fermi level is at zero energy.}
\label{fig:Fe_dos}
\end{figure}
 
In Figure \ref{fig:Fe_dos} we show the spin-resolved 3d density of states of bcc Fe for plain LDA, one-shot LDA+DMFT and CSC LDA+DMFT for U=1.7 eV and J=0.9 eV. Such a choice of U and J  was made considering the recently published LDA+DMFT study of the spin- and angular-resolved photoemission spectra of Fe (110) \cite{barriga09prl103:267203}, and the successful application of the LDA+DMFT scheme to the bcc-to-fcc ($\alpha$-$\gamma$) phase transition of Fe \cite{leonov11prl106:106405}. As observed in previous works \cite{grechnev07prb76:035107,katsnelson99jpcm11:1037,lichtenstein01prl87:067205} the local correlation effects are stronger for majority spin states, making the 3d band narrower, which is consistent with experiments \cite{barriga09prl103:267203}. Moreover the formation of a high energy satellite can be noticed, which has been thoroughly discussed in Ref. \onlinecite{grechnev07prb76:035107}. It is relevant to focus on the fact that essentially no differences can be observed between one-shot and CSC results for bcc Fe. In order to explore the role of the CSC cycle, we have made additional simulations for higher values of U (not shown here). Even for U as high as 4 eV no significant differences can be seen between one-shot and CSC densities of states. This is related to two factors. First of all our choice of the double counting cancels much of the effects of the Coulomb interaction, similarly to the so-called around-mean field double counting for LDA+U \cite{anisimov97jpcm9:767}. Additionally the fact that the 3d states only hybridize with very flat sp bands, makes the electron density insensitive to small changes induced by the self-energy.

\begin{table}
\caption{\label{tab:Fe}3d contribution to the occupation n, spin moment m$_\text{s}$, and orbital moment m$_\text{o}$ for bcc Fe. RSPt results for plain LDA, one-shot LDA+DMFT and CSC LDA+DMFT are presented for the two chosen values of U. All the magnetic moments are given in units of $\mu_B$ per atom.} 
\begin{tabular}{| l | c c c | c c c | }
\hline
 & \multicolumn{3}{c|}{one-shot} & \multicolumn{3}{c|}{CSC} \\
Method & n & m$_\text{s}$ & m$_\text{o}$ & n & m$_\text{s}$ & m$_\text{o}$ \\
\hline
LDA  &  6.24 & 2.21 &  0.05 &  6.24 & 2.21 &  0.05  \\
U=1.7 eV  & 6.27 & 2.17 &  0.07 &  6.25 & 2.14 &  0.06  \\
U=2.3 eV  & 6.28 & 2.19  & 0.06 & 6.25 &  2.15 & 0.06 \\
U=3.5 eV  & 6.32 & 2.25  & 0.05 & 6.27 &  2.22 & 0.04 \\
Exp. \cite{eriksson90prb42:2707,stearns:magnetostriction} & - & 2.13 & 0.08  & - & 2.13 & 0.08 \\
\hline
\end{tabular}
\vspace{0.2cm}

\end{table}

Despite the fact that the spectral properties look similar for the one-shot and the CSC calculations, differences can be observed when looking at the occupation numbers, spin and orbital moments, as show in Table \ref{tab:Fe}. The 3d occupation in one-shot LDA+DMFT tends to increase when compared to the bare LDA value, while the CSC calculation results in an occupation close to the LDA result. Such a behavior can be attributed to the fact that the $\op{H}^{LDA}_\bk$ is now fixed, and not allowed to adjust to compensate the changes induced by the local interactions in the electron density. Moreover from Table \ref{tab:Fe} we can also notice how the LDA+DMFT values improve on the LDA results for the spin and orbital moments, when compared to experimental data, especially for U=1.7 eV. However, the agreement is not perfect, and we identify the problem in the perturbative nature of the SPTF solver. Starting from the SPTF functional other computational schemes can be constructed, which do not suffer of such a problem \cite{sptf_igors_paper}.

\begin{figure}[t]
\centering
\includegraphics[scale=0.28]{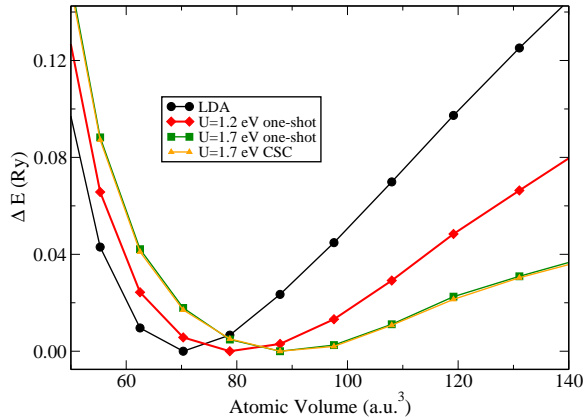}
\vspace{0.2cm}
\caption{(Color online) Total energy versus atomic volume curves of bcc Fe for LDA, one-shot LDA+DMFT and CSC LDA+DMFT for two different values of the local Coulomb interaction . The total energies are plotted with respect to their minima.}
\label{fig:Fe_energies}
\end{figure}

Although the differences of the occupation numbers due to the CSC cycle are small, we may expect more important changes to be seen in the total energy corrections to the bare LDA functional. Following the scheme presented in Ref. \onlinecite{dimarco09prb79:115111}, we calculated the LDA+DMFT total energy for a wide range of lattice constants. To reach a convergence up to the meV, a higher number of Matsubara frequencies was needed, i.e. 8192 frequencies, leading to an energy cut-off of about 20 Ry. The energy versus atomic volume curves for LDA and LDA+DMFT are shown in Figure \ref{fig:Fe_energies}. Already for such a small value of the Coulomb interaction as U=1.7 eV the changes on the ground state properties of the material are strong: the minimum is shifted to higher atomic volumes and the curve is flattened. Contrary to what one might expect, also for the total energies the differences between one-shot and CSC calculations are very small, and the corresponding curves in Figure \ref{fig:Fe_energies} are almost coinciding. For an ideal description of the ground-state properties of bcc Fe within our theory a smaller value of U is needed: the data for LDA+DMFT total energies for U=1.2 eV and J=0.8 eV are also shown in Figure \ref{fig:Fe_energies}. A better overview of these results can be obtained by looking at Table \ref{tab:Fe_energies}, where the equilibrium atomic volume and the bulk modulus are reported. Such values were obtained by fitting the total energy data through the Birch-Murnaghan equation of state\cite{murnaghan44,birch52}. Fitting through the Vinet equation of state\cite{vinet89jpcm1:1941} was also made, leading to very similar results (not shown here). As for our previous study of fcc Ni \cite{dimarco09prb79:115111}, it is clear that our calculations tend to over-estimate the role of the local exchange and correlation effects in the total energy. We identify two possible reasons for such a behavior. The first one can be found in the perturbative nature of the SPTF solver, which is supposed to be used for systems where the Hubbard U is smaller than the bandwidth W \cite{katsnelson02epjb30:9}. When increasing the lattice constant the bandwidth decreases, and the applicability of our approximations becomes questionable, at least concerning such a subtle feature as the total energy. The second reason is that the SPTF solver is usually applied in a non-conserving form, in Baym-Kadanoff sense \cite{sptf_igors_paper}, and this can introduce an additional error in the total energy. Although such a feature had already been observed for fcc Ni, it is especially interesting to see that it is present also for bcc Fe, where the bandwidth is bigger and the Coulomb repulsions are weaker. 

%

\begin{table}
\begin{tabular}{| l | c c | c | c c | c | }
\hline
 & \multicolumn{2}{c|}{LDA} & \multicolumn{1}{c|}{U=1.2 eV}& \multicolumn{2}{c|}{U=1.7 eV} & \multicolumn{1}{c|}{Exp.}\\
 & LMTO & LAPW & one-shot & one-shot & CSC &   \\
\hline
V$_0$ (a.u.$^3$)  & 70.49 & 70.71 & 78.51 & 87.00 & 87.06 & 79.51 \\
B (GPa) & 234 & 233 & 148 & 91 & 90 & 168 \\
\hline
\end{tabular}
\vspace{0.2cm}
\caption{\label{tab:Fe_energies} Equilibrium atomic volume $V_0$ and bulk modulus $B$ from LDA, one-shot LDA+DMFT and CSC LDA+DMFT for two different values of U. These values were obtained by fitting the calculated data through the Birch-Murnaghan equation of state\cite{murnaghan44,birch52} for a temperature $T=400 \text{K}$. Experimental data\cite{zeng08jpcm20:425217} and results of previous full potential LAPW studies\cite{kodera10jpsj79:074713} are shown for comparison.} 
\end{table}

\subsection{SmCo$_5$}
SmCo$_5$ has the CaCu$_5$ type crystal structure given by space group \#191 ($P6/mmm$) with lattice parameter $a = 5.002$ \AA{} and $c = 3.964$ \AA, with Sm in Wyckoff position 1a, and Co both in 2c and 3g. The {\bf{k}}-point mesh was set to $8 \times 8 \times 12$, giving 448 points in the irreducible wedge of the Brillioun zone. 

\begin{figure}
\includegraphics[scale=0.28]{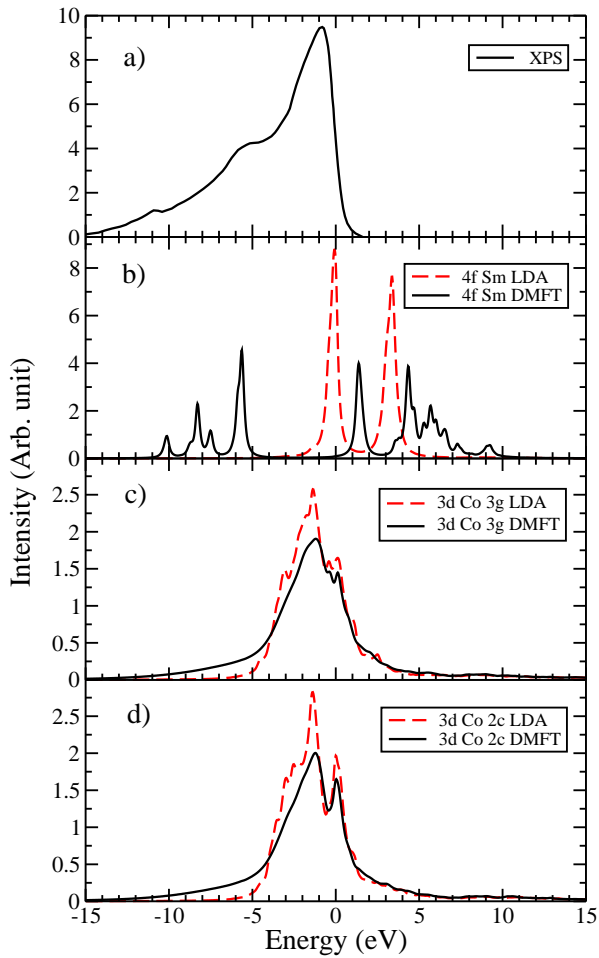}
\caption{(Color online) Experimental and calculated spectra of SmCo$_5$, with the Fermi level at zero energy. a) Experimental x-ray photoemission spectrum from Ref. \onlinecite{cuthill74aip18:1039}. The small peak at -11 eV is an Ar 2p artifact from the sample cleaning procedure. b) Calculated projected density of states of Sm 4f orbitals in LDA (dashed red line) and LDA+DMFT (thick black line). c) Calculated projected density of states of Co 3d orbitals at the Wyckoff position 3g in LDA (dashed red line) and LDA+DMFT (thick black line). d)  Calculated projected density of states of Co 3d orbitals at the Wyckoff position 2c in LDA (dashed red line) and LDA+DMFT (thick black line).}
\label{fig:pdos_vs_xps}
\end{figure}

In the LDA density of states, as shown in Fig. \ref{fig:pdos_vs_xps}b, the Sm 4f states are found in two narrow spin split peaks at 0 and 3.4 eV. The small bandwidth indicates that the f-electrons are strongly localized and should be treated as atomic-like. The Hubbard-I Approximation\cite{hubbard_I,lichtenstein98prb57:6884,svane06ssc140:364} (HIA) has been shown to give excellent results for strongly correlated f-orbitals\cite{lebegue05prb72:245102,lebegue06jpcm18:6329,thunstrom09prb79:165104} and was therefore chosen as the impurity solver for the Sm 4f orbitals. The Co 3d states form a broad peak between -4 to 0 eV in the LDA density of states, close to what is found in Co metal\cite{grechnev07prb76:035107,barriga10prb82:104414}, which points to weak correlation effects. With this in mind, the SPTF impurity solver in the fully renormalized version presented in Ref. \onlinecite{sptf_igors_paper}, was chosen to treat the Co 3d states.

The double counting correction $\op{H}_R^{DC} = \mu_{DC}\op{1} + \op{H}_X$ was used for the Sm 4f states, where the double counting parameter $\mu_{DC}$ acts as an atomic chemical potential, and $\op{H}_X$ was constructed to remove the local intra-orbital LDA exchange splitting. It is not possible to fully determine $\mu_{DC}$ from the number of Sm f-electrons alone, unless the system is intermediate valence\cite{thunstrom09prb79:165104}. In the present calculation the value of $\mu_{DC}$ was set by requiring that the Sm ground state should contain 5 electrons, and that the lowest binding energy of the $f^5$ configuration should be at 5.7 eV. $\op{H}_X$ was constructed under the assumption that the Sm 4f inter-orbital LDA exchange splitting was only slightly perturbed by the polarization of the f-states. An estimate of the inter-orbital LDA exchange splitting could then be obtained from a separate charge self-consistent calculation, but where the Sm 4f electrons were constrained to stay paramagnetic. One-shot calculations would completely neglect the effect of the SPTF self-energy on inter-orbital exchange splitting, and therefore charge self-consistence is necessary for an accurate calculation. The double-counting for the Co atoms is chosen to be the same as that for bcc Fe in the previous section.

The values of the Slater parameters F$^2$, F$^4$ and F$^6$ for the Sm 4f orbitals were recalculated at each LDA iteration from the radial integration of the bare Coulomb interaction over the f-orbitals. To account for some of the screening by the non-f-electrons, the values of the Slater parameters F$^2$ and F$^4$ were reduced to 92\% and 97\% of their calculated values, respectively. A reliable value of the Slater parameter F$^0$ can not be obtained in this way, as it is strongly affected by the screening. Therefore F$^0$ for the Sm 4f state was set to 8 eV, following Ref.\onlinecite{thunstrom09prb79:165104}. The Slater parameters for the Co 3d orbitals were taken from previous LDA+DMFT studies \cite{grechnev07prb76:035107}. A summary of the solvers, and the Coulomb parameters used in the LDA+DMFT calculation is given in Table \ref{tab:smco5params}.

\begin{table}
\caption{\label{tab:smco5params}Summary of the impurity solvers (AIS) and final Slater parameters (F$^0$,F$^2$, F$^4$ and F$^6$) used in the LDA+DMFT calculation of SmCo$_5$.}
\vspace{0.2cm}
\begin{tabular}{| l | c | c | c | c | c |}
\hline
Corr. orbitals & AIS & F$^0$ (eV) & F$^2$ (eV) & F$^4$ (eV) & F$^6$ (eV)\\
\hline
Sm 4f & HIA &  8.00 & 11.63 & 7.64 & 5.65 \\
Co 3d & SPTF & 2.50 & 7.75 & 4.85 & -- \\
\hline
\end{tabular}
\end{table}

The projected density of states of the Sm 4f and Co 3d orbitals from the CSC LDA+DMFT and LDA calculations are shown in Figure \ref{fig:pdos_vs_xps}. Both LDA and LDA+DMFT correctly position the Co 3d states at the Fermi level, but only the latter gets the shape of the peak in excellent agreement with the XPS spectrum of Ref. \onlinecite{cuthill74aip18:1039}. The differences are much larger for the Sm 4f states, where LDA yields two narrow spin-split peaks instead of the multiplet structures found in LDA+DMFT. The position of the first Sm 4f multiplet peak in the LDA+DMFT calculation at -5.7 eV coincides with the peak of the XPS spectrum, since the double counting parameter $\mu_{DC}$ was tuned with respect to this peakposition. The lack of structure between -10 to -7 eV in the experimental XPS spectrum is puzzling, as one would expect to see signs of the remaining multiplet structure like in Sm metal and SmCo$_2$\cite{kang93prb48:10327}.

 Finally, the magnetic moments are shown in Table V and the total moment from the theory is $10\%$ lower than the experimental data at zero Kelvin. Theoretical values obtained by LDA are also shown, but they are much smaller than in experiment.The data in Table V are rewarding in that for the Sm atom, the Russel-Saunders limit seems to be approaching, which is expected for a reasonable theoretical treatment of rare-earth based compounds.
\begin{table}
\caption{\label{tab:smco5mom}Spin (m$_\text{s}$), orbital (m$_\text{o}$), and total (m$_\text{tot}$) moment of SmCo$_5$ compared with experimental data at T=0 K. All the magnetic moments are given in units of $\mu_B$ per atom, while the final total moment is per formula unit. The contribution from the interstitial regions amount to 0.4 $\mu_B$. }
\vspace{0.2cm}
\begin{tabular}{| c | c c | c  c | c c | c |}
\hline
 & \multicolumn{2}{c|}{Sm} & \multicolumn{2}{c|}{Co (2c)} &  \multicolumn{2}{c|}{Co (3g)} & Total \\
Method & m$_\text{s}$ & m$_\text{o}$ & m$_\text{s}$ & m$_\text{o}$ &  m$_\text{s}$ & m$_\text{o}$ & m$_\text{tot}$ \\
\hline
LDA & -5.48 & 1.81 &  1.58 & 0.06 & 1.55 & 0.10 & 4.04  \\
LDA+DMFT &  -3.47 & 3.26 & 1.54 & 0.22 & 1.52 & 0.18 & 8.02 \\
~~~Exp.\cite{zhao91prb43:8593} & -- & -- & -- & -- & -- & -- & 8.9 \\
\hline
\end{tabular}
\end{table}

\section{Summary and conclusions}
We have developped a charge self-consistent LDA+DMFT code, and have applied it to the study of the electronic properties of a few weakly and strongly correlated materials. Antiferromagnetic NiO was used as a first test of our implementation, and comparison with other codes for electronic structure lead to a good agreeement, taking into account the technical differences involved in the calculations. Then we have addressed the problem of the role of the charge self-consistence in the LDA+DMFT study of the weakly correlated bcc Fe. It was found that the update of the LDA electronic density leads to minute changes of the spectral and magnetic properties. The total energy and the bulk modulus are also quite unchanged. However for the strongly correlated SmCo$_5$ the charge self-consistence is strictly needed to define a proper double-counting. Excellent agreement was found with existing photoemission data for low binding energies, but not for higher energies. In perspective we plan to apply the present methodology to study more complex properties of permanent magnets, complex oxides and corelated f-electron systems. Such studies are underway.

\section*{ACKNOWLEDGMENTS}
Financial support from the Swedish Research Council (VR) and Energimyndigheten (STEM), is acknowledged. Calculations have been performed at the Swedish national computer centers UPPMAX, PDC, HPC2N and NSC. O.E. is also grateful to the European Research Council (grant 247062 - ASD) and the Knut and Alice Wallenberg Foundation, for financial support.  

\appendix

\section{The full potential electron density}
\label{density}
In the following Appendix, we present some technical details of the calculation of the electron density from a density matrix in the FP-LMTO basis. The FP-LMTO basis is technically involved, and rather than going through all details, we will simply state the relevant expressions in the notation of Ref. \onlinecite{wills:fp-lmto} with only a brief explanation of the symbols.

In a normal single particle LDA calculation the density matrix  of Equation \eqref{eqn:rholdaan} is obtained by squaring the eigenvectors, weighing each vectors contribution by the appropriate Fermi weight,
\begin{equation}
\langle i | \op{\rho}_\bk^{LDA} | j \rangle \equiv \rho_{ij}(\bk) = \sum_v w_{v,\bk} \mA_i(v,\bk) \mA_j^{\dagger}(v,bk).
\label{appdensmat}
\end{equation}
Here $\mA_i(v,\bk)$ is basis function $i$ component of the eigenvector $v$, and $w_{v,\bk}$ is the corresponding Fermi weight.

\subsection{Within the muffin-tin spheres}

From a density matrix such as that given in Equation (\ref{appdensmat}), the electron density $n_t(\br)$ inside the muffin-tin of type $t$, is given as the following sum,
\begin{equation}\label{mtdens}
\begin{split}
n_{t}(r) = \sum_{h} D_{ht}(\rhat) & \sum_{e'\ell'} \tU_{t\ell'}(e', r) \\
& \sum_{e\ell}  M_{ht}(e, \ell; e', \ell') \tU^{T}_{t\ell}(e, r) ,
\end{split}
\end{equation}
where $\tU$ is a radial basis function and $D_{ht}$ is a symmetric linear combination of spherical harmonics.
We define the two intermediate density matrices,
\begin{widetext}
\begin{equation}
M_{ht}(e, \ell; e', \ell') = \frac{2\ell_h+1}{4\pi}\sum_{m, m', m_h} \mG(\ell, m; \ell', m'; \ell_h, m_h)
  \alpha^{*}(m_h)  \mM_t(e, \ell, m; e', \ell', m')
\end{equation}
\begin{equation}\label{eq:mldmtx}
\begin{split}
\mM_t(e, \ell, m; e', \ell', m') = \frac{1}{N_{\btau}(t)} \sum_{\btau\in t} \sum_{\bk}\sum_{i,j} \delta(e, e_{i})
& \Omega_{t,l}(e,\kappa_i)\tS_{\ell m;\ell_i m_i}(e; \kappa_{i}; \btau-\btau_{i}; \bk)
\times \\
& \rho_{ij}(\bk) \Omega_{t,l'}(e',\kappa_j)
  \tS^{\dagger}_{\ell' m';\ell_j m_j}(e'; \kappa_{j}; \btau-\btau_{j}; \bk)\delta(e', e_{j}),
\end{split}
\end{equation}
\end{widetext}
the first of which is given in terms of a Gaunt-like coefficient, $\mG$, and a projection coefficient onto a harmonic function, $\alpha_{ht}$. In the second equation, $\rho_{ij}$ is multiplied by the LMTO structure constants, $\tS$, rescaled by the matching coefficient matrix, $\Omega$, that matches the atomic solutions inside the muffin tin to the Bessel and Neumann functions in the interstitial region (as well as their respective derivatives).
The quantities $\mM_t$ and $M_{ht}$ describe the partial occupancies inside a muffin tin, and we may also use them to calculate, for example, the total orbital moment,
\begin{equation}
\begin{split}
M = \sum_{e,e',\ell,m} m \cdot & \mM_t(e,\ell,m,e',\ell',m') \\
 & \mO_t(e,e',\ell,\ell')\delta(\ell, \ell')\delta(m,m'),
\end{split}
\end{equation}
or, by summing away the energy set indices, we can get the total $\ell$-charge,
\begin{equation}
\begin{split}
N_{\ell} = \sum_{e,e',m} & \mM_{t}(e,\ell,m,e',\ell',m') \\
& \mO_t(e,e',\ell,\ell')\delta(\ell, \ell')\delta(m,m') .
\end{split}
\end{equation}
Here we also need to account for $\mO_t(e,e',\ell,\ell')$, the overlap between the radial basis functions on type $t$.

\subsection{In the intersititial region}

The density in the interstitial is constructed from the $\bk$-dependent density matrix $\rho_{ij}(\bk)$ from Equation (\ref{appdensmat})
and {\it pseudo}basis functions $\tilde{\psi}_{i}(\bk)$, which are built with the constraints of matching the true basis functions in the 
interstitial and having a rapidly convergent Fourier representation.  The $\tilde{\psi}$ are, in practice,
defined by their Fourier representation, and we have 
\begin{eqnarray}
      \tilde{n}(r)     &=& \sum_{k} n(\bk,r) 
\nonumber \\
      \tilde{n}(\bk,r) &=& \Tr \sum_{ij} \tilde{\psi}_{i}(\bk,r) \rho_{ij}(\bk) 
                                         \tilde{\psi}^{\dagger}_{j}(\bk,r) 
 \\ \nonumber
               &=& \Tr \sum_{ij} F^{-1}[\tilde{\psi}_{i}(\bk,g)] \rho_{ij}(\bk) 
                                 F^{-1}[\tilde{\psi}^{\dagger}_{j}(\bk,g)] 
\label{appendin1}\end{eqnarray}
where $F^{-1}$ is the inverse Fourier transform. 
The pseudo-density $\tilde{n}(r)$ is identical to the true density $n(r)$ in the interstitial.  
The construction of $\tilde{\psi}$ is described in Ref. \onlinecite{wills:fp-lmto}.

Because the density matrix is Hermitian, the density is accumulated for each $\bk$ by 
\begin{eqnarray}
      \tilde{n}(\bk,r) &=& \Tr \sum_{i} F^{-1}[\tilde{\psi}_{i}(\bk,g)]  \Big(\sum_{j<i}2\rho_{ij}(\bk) F^{-1}[\tilde{\psi}^{\dagger}_{j}(\bk,g)] \nonumber  \\
                   &&  + \rho_{ii}(\bk) F^{-1}[\tilde{\psi}^{\dagger}_{i}(\bk,g)]
                   \Big)        .            
\label{appendin2}\end{eqnarray}

\bibliography{strings,kylie}

\end{document}